\documentstyle[aps,times,epsf,12pt]{revtex}
\newcommand{\tfrac}[2]{{\textstyle\frac{#1}{#2}}}
\begin{document}

\draft

\title{Phases of a Stack of Membranes at Large-$\symbol{100}$} 

\author{M.E.S. Borelli \thanks{E-mail: borelli@physik.fu-berlin.de}
\and H. Kleinert
\thanks{E-mail:kleinert@physik.fu-berlin.de}\address{Institut f\"ur
Theoretische Physik \\ Freie Universit\"at Berlin \\ Arnimallee 14,
14195 Berlin, Germany}}
\date{\today}

\maketitle

\begin{abstract} The phase diagram of a stack of tensionless membranes with
nonlinear curvature energy and vertical harmonic interaction is
calculated exactly in a large number of dimensions of configuration
space. At low temperatures, the system forms a lamellar phase with
spontaneously broken translational symmetry in the vertical direction.
At a critical temperature, the stack disorders vertically in a
melting-like transition.  The critical temperature is determined as a
function of the interlayer separation $l$.
\end{abstract}

\pacs{{\it PACS}: 82.65.Dp, 68.35.Rh \\ {\it Keywords}: Thermodynamics
of surfaces and interfaces; Phase transitions and critical phenomena}

\section{Introduction} \label{intro}

Recently, a model for a finite stack of tensionless membranes
\cite{helfstack} was studied with respect to the effects of
higher-order terms of the curvature energy \cite{stack}.  The approach
was perturbative, using the renormalization group to sum infinitely
many terms.  It was shown that thermal fluctuations induce the melting
of the stack into a vertically disordered phase. By its nature, the
perturbative expansion was able to give a satisfactory description only
for the ordered phase.

For a description of the disordered phase and a better understanding
of the entire transition, we analyze in this paper the behavior of a
stack of tensionless membranes exactly for very large dimension $d$ of
the embedding space.  Since the model is exactly solvable in this
limit, we can calculate all its relevant properties explicitly, in
particular its complete phase diagram as a function of the interlayer
separation $l$.

\section{The Model} \label{model}

As in Ref.\ \cite{stack}, we consider a model in which a multilayer
system is made up of $(N+1)$ fluid membranes, parallel to the
$xy$-plane of a Cartesian coordinate system, separated a distance $l$.
If the vertical displacement of the $m$th membrane with respect to
this reference plane is described by a function $u_m({\bf x}) \equiv
u({\bf x}_\perp,m l)$, where ${\bf x}_\perp = (x,y)$, the energy of
the stack reads:
\begin{equation} \label{mod0}
E = \sum_{m} \int {\rm d}^2 x_\perp \sqrt{g_m} \left[\tfrac{1}{2}
\kappa_0 H_m^2 +  \frac{B_0}{2 l} (u_m - u_{m-1})^2 \right].
\end{equation}
Here, $H_m= \partial_a N_{m,i}$ is the mean curvature, where ${\bf
N}_m \propto (-\partial_1 u_m, -\partial_2 u_m, 1)$ is the unit normal
to the $m$th membrane, and 
\begin{equation}
  \label{gmunu}
	g_{m,i j}=\delta_{i j} + \partial_i u_m\partial_j u_m
\end{equation} 
the induced me\-tric, with $i,j = 1,2$, $\partial_1 =
\partial/\partial x, \partial_2 = \partial/\partial y$ and $g_m = \det
[g_{m,i j}]$. The parameter $\kappa_0$ is the ben\-ding rigidity of a
single membrane, and $B_0$ the compressibility of the stack.  In
Eq.\ (\ref{mod0}), as in the following, the subscript $0$ denotes bare
quantities, whereas renormalized parameters will carry no subscript.

For slow spatial variations, the discrete variable $m l$ may be replaced
with a continuous one, and $u({\bf x}_\perp,m l) \to u({\bf x})$, where
${\bf x} = ({\bf x}_\perp,z)$.  In this limit, the ener\-gy
(\ref{mod0}) reduces to
\begin{equation} \label{mod1}
E = \int^{L_{\parallel}}_0 
{\rm d}z \int {\rm d}^2 x_\perp \sqrt{g} \left[
\tfrac{1}{2} K _0 H^2 + \tfrac{1}{2} B_0 (\partial_z u)^2 \right].
\end{equation}
Here we have introduced the bulk version of the ben\-ding rigidity
$K_0 \equiv \kappa_0/l$, and defined $L_{\parallel} \equiv N l$.  The
gradient energy $(\partial_z u)^2$ has the unphysical feature that it
gives $z$-dependent reparametrizations a kinetic energy.  It should
therefore be replaced by the normal gradient energy $({\bf N} \cdot
\nabla u)^2$.  We have seen in Ref.\ \cite{stack} that this changes
the critical exponent with which the renormalized $B$ vanishes as the
bending rigidity becomes critical.  However, in the limit of large
$d$, which we are going to investigate, the difference between the two
gradient energies will be negligible.

\section{Large-$\symbol{100}$ analysis} \label{larged}

For arbitrary $d$, the vertical displacement of the $m$th membrane in
the stack becomes a $(d-2)$-vector field ${\bf u}_m({\bf x}_\perp)$.

\subsection{Partition function and the energy}

It is useful to consider $g_{i j}$ as an independent field
\cite{polyakov}, and impose relation (\ref{gmunu}) with help of a
Lagrange multiplier $\lambda_{i j}$.  We write the partition function
as a functional integral over all possible configurations ${\bf
u}_m({\bf x}_\perp)$ of the individual membranes in the stack, as well
as over all possible metrics $g_{m,i j}$.  After taking again the
continuum limit, the partition function reads:
\begin{equation} \label{partfunc3}
Z = \int {\cal D}g \, {\cal D} \lambda \, {\cal D} {{\bf u}} \, {\rm
e}^{-E_0/k_{\rm B} T},
\end{equation}
with 
\begin{eqnarray} \label{action2}
&& E_0 = \int {\rm d} z \, {\rm d}^2 x_\perp \sqrt{g} \left\{  \sigma_0 +
\tfrac{1}{2} B_0 (\partial_z {\bf u})^2 + \tfrac{1}{2} K_0 (\partial_k^2
{\bf u})^2  + \tfrac{1}{2} K_0 \lambda^{i j}(\delta_{i j} +
\partial_{i} {\bf u} \partial_{j}{\bf u} - g_{i j}) - \tfrac{1}{4}
\tau_0 \lambda_{i i}^2 \right\}, \nonumber \\ &&
\end{eqnarray}
where ${\bf u}$ is a $(d-2)$-dimensional vector-function of ${\bf
x}_\perp, z$.  Note that the functional integral over $\lambda$ in
(\ref{partfunc3}) has to be performed along the imaginary axis to
result in a $\delta$-function.  We have also introduced a term
proportional to $\lambda_{i i}^2$.  This term is necessary to
renormalize the theory, and its coefficient $\tau_0$ corresponds to
the large-$d$ in-plane compressibility of the membranes.  Since we
take the membranes to be incompressible, we shall set the renormalized
$\tau$ equal to zero at the end of our calculations. We have also
included a surface tension $\sigma_0$, again to absorb infinities and
to be set equal to zero after renormalization.

The functional integral over ${\bf u}$ in Eq.\ (\ref{partfunc3}) is Gaussian
and can be carried out to yield an effective energy
\begin{equation} \label{effectaction}
E_{\rm eff} = \tilde{E}_0 + E_1, 
\end{equation}
with
\begin{equation}
\tilde{E}_0 = \int {\rm d} z \, {\rm d}^2 x_\perp \sqrt{g} \left[ \sigma_0 +
\tfrac{1}{2} K_0 \lambda^{i j}(\delta_{i j} - g_{i j}) - \tfrac{1}{4}
\tau_0 \lambda_{i i}^2 \right]
\label{s0tilde},
\end{equation}
and
\begin{equation}
E_1 = \frac{d-2}{2} k_{\rm B} T \, \mbox{Tr} \ln \left[ B_0
\omega^2 + K_0( q_\perp^4 - q_i \lambda^{i j} q_j)
\right], \label{s1}
\end{equation}
where the functional trace Tr is here an integral over space as well
as the integral over wavevectors ${\bf q}_\perp$ and $\omega$, after
replacing $\partial_z^2 \to - \omega^2$ and ${g^{i j}\partial_{i}
\partial_{i} \to - \bf q}^2$.  Note that the discrete nature of the
stack restricts the integral over the wavevectors $\omega$ to the
first Brillouin zone $|\omega| < \pi/l$.

In the large-$d$ limit, the partition function (\ref{partfunc3}) is
do\-minated by the saddle point of the effective energy
(\ref{effectaction}) with respect to the metric $g_{i j}$ and the
Lagrange multiplier $\lambda^{i j}$.  For very large membranes, the
saddle point can be assumed to be symmetric and homogeneous
\cite{kleinld,olesen,david}:
\begin{equation}
g_{i j}= \varrho_0 \delta_{i j}; \; \lambda^{i j} = \lambda_0 g^{i j} =
\frac{\lambda_0}{\varrho_0} \delta^{i j},
\end{equation}  
with constant $\varrho_0$ and $\lambda_0$. At the saddle point the
effective energy (\ref{effectaction}) becomes the free energy of the
system.

In the following we shall investigate both the case of an infinite and a
finite stack of membranes.  As we will see, the large-$d$ approximation allows
for the vertical melting even in an infinite stack, which is not found
perturbatively \cite{helfstack}.

\subsection{Infinite stack} \label{infinite}

Let us first analyze the case of an infinite stack.  To simplify our
calculations, we assume the number $N+1$ of membranes in the stack to
be very large, making the distance $l$ between them very small.  In this
regime, we may extend the limits $\pm \pi/l$ of the integral over
$\omega$ to infinity.  The explicit $l$-dependence will be introduced
later into our calculations.

After evaluating the functional trace in Eq.\ (\ref{s1}), we obtain
\begin{equation} \label{s1linf}
E_1 = \frac{d k_{\rm B} T}{2}  \int {\rm d}z {\rm d}^2 x_{\perp}
 \varrho_0
  \sqrt{\frac{K_0}{B_0}} \left\{ \frac{\Lambda^4}{8 \pi}  +
  \frac{\lambda_0}{8 \pi} \Lambda^2  
+ \frac{\lambda_0^2}{64 \pi} \left[ 1 - 2
  \ln \left( \frac{4 \Lambda^2}{\lambda_0} \right)\right]\right\},
\end{equation}
where ultraviolet divergences are regularized by introdu\-cing a sharp
transverse wavevector cutoff $\Lambda$ and $d-2$ has been replaced by
$d$ for large $d$.

We may now absorb the first term in (\ref{s1linf}) by renormalizing
$\sigma_0$, so that
\begin{equation}
\sigma = \sigma_0 + \frac{d k_{\rm B} T }{16 \pi} \sqrt{\frac{K_0}{B_0}} \Lambda^4 
\end{equation}
is the physical surface tension, which is set equal to zero.  The
second, quadratically divergent term in (\ref{s1linf}) is used to
define the critical temperature as 
\begin{equation}
  \label{Tclinf}
  \frac{1}{T_{\rm c}} \equiv \frac{d k_{\rm B}}{16 \pi}
  \frac{\Lambda^2}{\sqrt{B K}} .
\end{equation}
The next divergent term, proportional to $\lambda_0^2$, is regularized
by introducing a renormalization scale $\mu$ and modifying the the
in-plane compressibility to
\begin{equation}
  \label{compressibility}
  \tau = \tau_0 + \frac{d k_{\rm B} T}{32 \pi} 
  \sqrt{\frac{K_0}{B_0}} \ln \left( 4 e^{-1/2}
    \frac{\Lambda^2}{\mu^2} \right).
\end{equation}
The physical in-plane compressibility $\tau$ is now set equal to zero,
as explained in the previous section.

The effective energy thus becomes
\begin{equation}\label{Eefflinf}
  E_{\rm eff} = \int {\rm d}z \, {\rm d}^2 x_{\perp} K \lambda \varrho \left\{
  \left( \frac{1}{\varrho} -1 \right) + \frac{T}{T_{\rm c}}  
  +  \frac{a T}{\sqrt{K}}
  \lambda \left[ \ln \left( \frac{\lambda}{\bar{\lambda}}\right) -
    \frac{1}{2}\right] \right\},
\end{equation}
with the constants $a \equiv  d k_{\rm B} /64 \pi \sqrt{B}$, $\bar{\lambda}
\equiv \mu^2 e^{-1/2}$.

From the second derivative matrix of $E_{\rm eff}$
with respect to $\varrho$ and $\lambda$ we find that the stability of
the saddle point is guaranteed only for $\lambda < \bar{\lambda}$.

Extremizing the above expression with respect to $\varrho$, we find
two solutions for $\lambda$, namely $\lambda = 0$ and
$\lambda=\lambda_\infty$, with
\begin{equation}
  \label{eq:lambdalinf}
  \lambda_\infty \left[ \ln \left( \frac{\lambda_\infty}{\bar{\lambda}} \right) -
  \frac{1}{2} \right] = \frac{\sqrt{K}}{a} \left( \frac{1}{T} - \frac{1}{T_{\rm
  c}} \right).
\end{equation}

For $T < T_{\rm c}$, this equation has no solution for
$\lambda_\infty$.  In this case, the only possible solution is
$\lambda = 0$, which corresponds to the ordered phase as we shall
verify later.  For $T > T_{\rm c}$, the saddle point lies at $\lambda
= \lambda_\infty$, which is now well-defined.  This is the vertically
disordered phase.

\begin{figure}[h]
\begin{center}
\epsfxsize=7.cm \mbox{\epsfbox{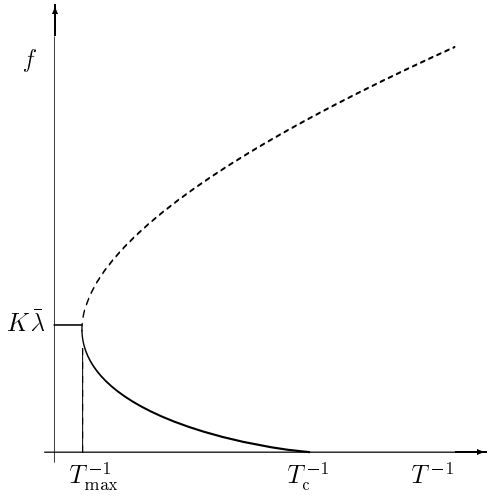}}
\end{center}
\caption{Free energy density of an infinite stack of
membranes.  The dashed curve indicates the unphysical branch of the
solution of Eq.\ (\ref{eq:lambdalinf}).}
\label{fig:lambda}
\end{figure}

The free energy density at the extremum is given by
\begin{equation}
  \label{eq:freeextrem}
  f = K \lambda_\infty 
\end{equation}
and its behavior is similar to the one found perturbatively in Ref.\
\cite{stack}. (see Fig.\ \ref{fig:lambda}).

Extremizing the effective energy (\ref{Eefflinf}) with respect to
$\lambda$, we find $\varrho$ as a function of temperature.  For $T <
T_{\rm c}$ it is given by
\begin{equation}
  \label{rho1linf}
  \varrho_{-}^{-1} = 1 -  \frac{T}{T_{\rm c}}.
\end{equation}
This as $T$
approaches $T_{\rm c}$ from below, indicating the vertical melting at
$T_{\rm c}$.  In the disordered phase, $\varrho$ is found to be
\begin{equation}
  \label{rho2linf}
  \varrho_{+}^{-1} =  \frac{T}{T_{\rm
  c}}  - 1 - \frac{a \lambda_\infty T}{\sqrt{K}}.
\end{equation}
As $T$ approaches $T_{\rm c}$ from above, $\lambda_\infty$ tends to
zero, and $\varrho$ goes again to infinity.

The positivity of $\varrho$ and the stability of the saddle point
imply that there is a maximum temperature, given by
\begin{equation}
\frac{1}{T_{\rm max}} = \frac{1}{T_{\rm c}} - \frac{a
\bar{\lambda}}{\sqrt{K}},
\end{equation}
below which our assumption that the membranes in the stack are
in-plane incompressible does not lead to a stable system.

\subsection{Finite stack of many membranes} \label{finite}  

Let us now analyze the case of a finite stack of size
$L_{\parallel}$. Now the functional trace in (\ref{s1}) involves a
sum over the discrete wavevectors $\omega_n$, given
by
\begin{equation} \label{matsubara}
\omega_n = \frac{2 \pi}{L_{\parallel}} n, \;\;\; n=0,\pm 1, \pm 2 , \cdots.
\end{equation}
For small $\lambda_0$, a series expansion leads to
\begin{equation}
  \label{s1finl}
  E_1 = \int {\rm d}z \, {\rm d}^2 \sigma \frac{d k_{\rm B} T}{2} \varrho_0 e_1,
\end{equation}
with
\begin{eqnarray}
  e_1 =  
  &&
  \frac{\Lambda^4}{8 \pi} \sqrt{\frac{K_0}{B_0}}   -  \frac{\pi}{12} \sqrt{\frac{B_0}{K_0}}\frac{1}{L_\parallel^2} 
  + \frac{\lambda_0}{8 \pi} \sqrt{\frac{K_0}{B_0}} \Lambda^2 
  + \frac{\lambda_0}{4 \pi L_\parallel} \ln
  \left( \frac{L_\perp^2}{L_\parallel} \sqrt{\frac{B_0}{K_0}}\right) \nonumber \\ &&
  + \frac{\lambda_0^2}{64 \pi} \sqrt{\frac{K_0}{B_0}} \left[3 - 2 \gamma + 2
    \ln \left( \frac{\lambda_0}{8 \pi \Lambda^2} \frac{L_\perp^2}{L_{\parallel}}
      \sqrt{\frac{B_0}{K_0}} \right) \right] \nonumber \\ &&  
  + \sqrt{\pi} \sum_{m=3}^{\infty}
  \frac{(-1)^{m+1}\lambda_0^m}{m \, 2^{2 m} \pi^m} L_\parallel^{m-2} 
  \left( \frac{K_0}{B_0}\right)^{\frac{m-1}{2}}
  \frac{\Gamma(\frac{m-1}{2})}{\Gamma(\frac{m}{2})} \zeta(m-1).
\end{eqnarray}
As in the case of the infinite stack, we absorb the logarithmic
divergence by renormalizing the in plane compressibility via Eq.\
(\ref{compressibility}), setting $\tau$ equal to zero for incompressible
membranes.  The surface tension receives now an $L_\parallel$-dependent
renormalization
\begin{equation}
  \sigma = \sigma_0 + \frac{d k_{\rm B} T }{16 \pi} \sqrt{\frac{K_0}{B_0}}
  \Lambda^4 -  d k_{\rm B} T \frac{\pi}{24} \sqrt{\frac{B_0}{K_0}} 
  \frac{1}{L_\parallel^2},
\end{equation}
and $\sigma$ is again set equal to zero to describe a stack of tensionless
membranes.

Extremization of the renormalized combined effective action
 (\ref{s0tilde}) and (\ref{s1finl}) with respect to $\varrho$ leads
 again to two possible solutions for the saddle point, namely $
 \lambda = 0$ or $\lambda = \lambda_{L_\parallel}$, with
\begin{eqnarray}
  \label{lambdafinl}
&& \lambda_{L_\parallel} \left[ \ln \left( 
\frac{\lambda_{L_\parallel}}{\bar{\lambda}} \right) -
  \frac{1}{2} \right] + \lambda_{L_\parallel} \left[ 1 - \gamma + \ln
  \left(\frac{L_{\perp}^2}{L_\parallel} \frac{1}{8 \pi} \sqrt{\frac{B}{K}}
  \right)\right]
  \nonumber \\ &&+ 32 \pi^{3/2} \sum_{m=3}^{\infty} \frac{ (-1)^{m+1} \lambda_{L_\parallel}^{m-1}}{m \, 2^{2 m} \pi^m} L_{\parallel}^{m-2} \left( \frac{K}{B}\right)^{\frac{m-2}{2}}
  \frac{\Gamma(\frac{m-1}{2})}{\Gamma(\frac{m}{2})} 
   \zeta(m-1) = \frac{\sqrt{K}}{a}  \left( \frac{1}{T} -
  \frac{1}{T_{L_\parallel}} \right)
\end{eqnarray}
where
\begin{equation}
  \label{Tcfinl} 
  \frac{1}{T_{L_\parallel}} =
  \frac{1}{ T_{\rm c}} +
  \frac{d k_{\rm B} }{8 \pi} \frac{1}{K L_{\parallel}} \ln
  \left(\frac{L_\perp^2}{L_\parallel} \sqrt{\frac{B}{K}}
  \right)
\end{equation}
is the inverse critical temperature for a stack of size $L_\parallel$ .

For $T < T_{L_\parallel}$, Eq.\ (\ref{lambdafinl})
has no solution. In this case, the stack is
in the ordered phase, the only available solution for the saddle
point being $\lambda = 0$.  For $T > T_{L_\parallel}$, there exists a nonzero
solution $\lambda_{{L_\parallel}}$, where the system is in the vertically
disordered phase.

Let us now examine the saddle point solutions for $\varrho$. In the vertically
disordered phase where $\lambda = \lambda_{L_\parallel}$ is nonzero, we may
expand the effective energy in a small-$L_{\parallel}$ series.  Extremization
with respect to $\lambda_{L_\parallel}$ leads to
\begin{eqnarray} 
  \label{rhoplus}
  \varrho_{+}^{-1} =&& \frac{T}{T_{L_\parallel}} - 1 -  \frac{a
    \lambda_{L_\parallel} T}{\sqrt{K}}
  \nonumber \\ &&
  - \frac{d k_{\rm B} T}{2 K} \sqrt{\pi}
    \sum_{m=3}^{\infty} \frac{(-1)^{m+1}\lambda_{L_\parallel}^{m-1}}{2^{2 m}
      \pi^m} \left(1 - \frac{2}{m} \right) {L_\parallel}^{m-2} 
    \left( \frac{K}{B}\right)^{\frac{m-1}{2}}
    \frac{\Gamma(\frac{m-1}{2})}{\Gamma(\frac{m}{2})} \zeta(m-1).
  \end{eqnarray}
  The positivity of $\varrho$ and the stability of the
  saddle point again define a maximal temperature,
given by
\begin{equation}
\frac{1}{T_{\rm max}^{L_{\parallel}}} =  \frac{1}{T_{\rm max}} -
 \frac{d k_{\rm B} T}{16 \pi K L_{\parallel}} \ln \left( \frac{16
 \pi}{d k_{\rm B} T_{\rm c}} \frac{\sqrt{B K}}{\bar{\lambda}}\right),  
\end{equation}
above which our assumption that the membranes in the stack are
in-plane incompressible cannot be maintained.

\begin{figure}[h]
\begin{center}
\epsfxsize=8.cm \mbox{\epsfbox{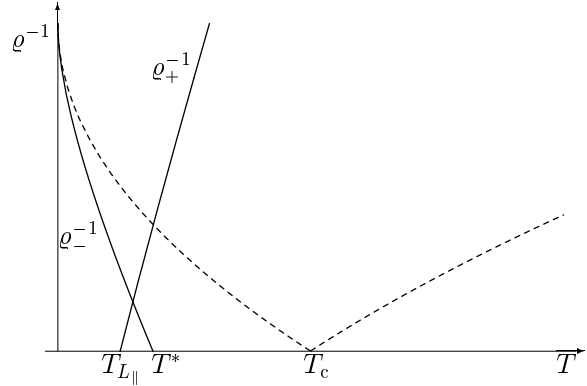}}
\end{center}
\caption{Behavior of $\varrho^{-1}$ as a function of $T$. The solid
lines indicate the solutions of the saddle point for $\varrho^{-1}$
for a finite stack. Above $T_{L_\parallel}$, $\varrho^{-1}$ is given
by (\ref{rhoplus}), and below $T_{L_\parallel}$ by (\ref{rhoftl0}).
The dashed lines indicate the behavior of $\varrho^{-1}$ for an
infinite stack.}
\label{fig:rho}
\end{figure}

In the ordered phase, the situation is more delicate.  For
$\lambda = 0$, $\varrho$ can be calculated exactly,
and we obtain
\begin{equation} \label{rhoftl0}
\varrho_{-}^{-1} = 1 - \frac{d k_{\rm B} T}{8 \pi K L_{\parallel}} \ln
\left[ \frac{\sinh \left( \frac{8 \pi K L_{\parallel}}{d k_{\rm B}
T_{\rm c}} \right)}{\frac{L_\parallel}{ 2 L_{\perp}^2}
\sqrt{\frac{K}{B}}}\right],
\end{equation}
with an infrared regulator $L_\perp$ equal to the inverse lateral size
of the membranes in the stack. If the size $L_{\parallel}$ of the
stack is large, (\ref{rhoftl0}) may be approximated by
\begin{equation}
\label{rhominus}
\varrho_{-}^{-1} \approx  1 - \frac{T}{T_{L_\parallel}}.
\end{equation}
For smaller stacks, however, the positivity of $\varrho$ is not
guaranteed. For a fixed, but small stack size $L_{\parallel}$, and for fixed
lateral size $L_\perp$ of the membranes in the stack, 
there is a characteristic temperature defined by
\begin{equation}
\label{Tcr}
T^* = \frac{8 \pi K L_{\parallel}}{d k_{\rm B}  \ln(16
\pi \sqrt{B K} L_{\perp}^2 / d k_{\rm B} T_{\rm c}) },
\end{equation}
above which $\varrho$ changes sign, and (\ref{rhoftl0}) is no longer
applicable.  Interestingly, for all $L_\perp$ and for all finite sizes
$L_\parallel$ of the stack, the
critical temperature $T_{L_\parallel}$ is lower than $T^*$, so
that the vertical melting still occurs.  The behavior of $\varrho$ is
depicted in Fig.\ \ref{fig:rho}.

Note that Eq.\ (\ref{Tcr}) reflects the existence of a characteristic
horizontal length scale. At fixed temperature $ T_{L_{\parallel}}< T <
T^*$, and for membranes of lateral size $L_{\perp}$ smaller than
\begin{equation}
  \label{Lp}
L_{\rm p} = \Lambda^{-1} \exp\left( \frac{4 \pi K L_\parallel}{d k_{\rm B} T  }\right),
\end{equation}
the height fluctuations of the individual membranes are not strong
enough to destroy the ordered phase.  The characteristic length
$L_{\rm p}$ corresponds to the the de Gennes-Taupin persistence length
$\xi_{\rm p}$ \cite{DGT} of the individual membranes, below which
crumpled membranes appear flat.

\subsection{Finite number of membranes}

Until now we have performed our calculations in the somewhat
unphysical continuum approximation, by letting the interlayer
separation $l$ be very small making the number of membranes in the
stack very large.  Let us now investigate the properties of the stack
for a fixed number $N+1$ of membranes at a finite interlayer distance
$l$.

For this purpose, we replace the continuum derivative $\partial_z^2$
in the $z$-direction with the discrete gradient operator $\nabla^2$,
whose eigenvalues are given by
\begin{equation}
  \label{eq:nabla}
  \nabla^2 g = \frac{2(1 - \cos \omega_n l)}{l^2} g,	
\end{equation}
where $g$ is some test function.  The discrete wavevectors $\omega_n$
are now given by
\begin{equation}
  \label{eq:omega_dirichlet}
  \omega_n = \frac{n \pi}{N l}, \; n= 1, 2, \cdots, N.
\end{equation}
For small interlayer separation $l$ , the free energy is given by
(\ref{s1finl}) with
\begin{eqnarray}
  e_1 &=& \frac{\Lambda^4}{8 \pi} \sqrt{\frac{K_0}{B_0}} + \frac{1}{N
  l^2}\sqrt{\frac{B_0}{K_0}} + \frac{\lambda_0}{8 \pi}
  \sqrt{\frac{K_0}{B_0}} \Lambda^2  +
  \frac{\lambda_0}{4 \pi N l} \left[ - \ln N + 2 N \ln \left( l
  \sqrt{\frac{K_0}{B_0}} \Lambda^2 \right) \right] \nonumber \\ &+&
  \frac{\lambda_0^2}{64 \pi} \sqrt{\frac{K_0}{B_0}} \left[ 1 - 2 \ln \left( \frac{4
  \Lambda^2}{\lambda_0} \right)\right] + \frac{1}{N
  \sqrt{\pi}} \sum_{m=2}^{\infty} \frac{(-1)^{m+1}\lambda_0^m}{m \,
  2^{\frac{3 m+1}{2}}} l^{m-2} \left(
  \frac{K_0}{B_0}\right)^{\frac{m-1}{2}} 
  \frac{\Gamma(\frac{m-1}{2})}{\Gamma(\frac{m}{2})}
  \bar{\zeta}_{N}(m-1), \nonumber \\
\end{eqnarray}
where we have defined the modified Zeta-function 
\begin{equation}
  \label{eq:zetan}
  \bar{\zeta}_{N}(m) = \sum_{n=1}^N \frac{1}{\left[ 1 - \cos\left(
  \frac{n \pi}{N}\right)\right]^\frac{m}{2}}. 
\end{equation}

We proceed by renormalizing the in-plane compressibility via Eq.\
(\ref{compressibility}), setting $\tau$ equal to zero for incompressible
membranes as before.  The surface tension receives an $l$-dependent
renormalization
\begin{equation}
  \sigma = \sigma_0 + \frac{d k_{\rm B} T }{16 \pi} \sqrt{\frac{K_0}{B_0}}
  \Lambda^4 -  \frac{d k_{\rm B} T} {2 N l^2} \sqrt{\frac{B_0}{K_0}} 
\end{equation}
and $\sigma$ is set equal to zero to describe a stack of tensionless
membranes, as before. But now the bulk bending rigidity is also
modified to
\begin{equation}
  \label{eq:kappa}
  K = K_0 - \frac{d k_{\rm B} T}{4 \pi l} \ln \frac{\Lambda^2}{\mu^2}.
\end{equation}
Note that this renormalization agrees with the known result for a
single membrane \cite{peliti}.

The saddle point for $\lambda$ is now given by
$\lambda = 0$ or $\lambda = \lambda_l$, with
\begin{eqnarray}
  \label{lambdadiscl} && \lambda_{l} \left[ \ln \left(
\frac{\lambda_l}{\bar{\lambda}} \right) - \frac{1}{2} \right] + 32
\sqrt{\pi} \sum_{m=2}^{\infty} \frac{ (-1)^{m+1} \lambda_{l}^{m-1}}{m
\, 2^{\frac{3 m + 1}{2}}} l^{m-2} \left(
\frac{K}{B}\right)^{\frac{m-2}{2}}\frac{\Gamma(\frac{m-1}{2})}{\Gamma(\frac{m}{2})}
\bar{\zeta}_N(m-1) \nonumber \\ && = \frac{\sqrt{K}}{a} \left(
\frac{1}{T} - \frac{1}{T_l} \right)
\end{eqnarray}
where
\begin{equation}
  \label{Tcdisc} 
  \frac{1}{T_l} = \frac{1}{ T_{\rm c}} +
  \frac{d k_{\rm B} }{8 \pi N \kappa } \left\{ -\ln N + 2 N \left[\ln
  \left(l \sqrt{\frac{K}{B}} \bar{\lambda}\right) + \frac{1}{2}
  \right] \right\}
\end{equation}
is the inverse $l$-dependent critical temperature, and $\kappa$ is the
bending rigidity of a single membrane in the stack.  The two solutions
for $\lambda$ again imply the existence of two different phases, with
a phase transition at the critical temperature $T_l$, which, as in the
perturbative case, depends only weakly on the number of membranes in
the stack.  The corresponding solutions for $\varrho$, obtained by
extremizing the effective energy with respect to $\lambda$, are given
by
\begin{equation}
  \label{rhominusfinl}
  \varrho_{-}^{-1} =  1 - \frac{T}{T_l},	
\end{equation}
for $T < T_l$, that is, in the ordered phase, and
\begin{eqnarray} 
  \label{rhoplusfinl} \varrho_{+}^{-1} =&& \frac{T}{T_l} - 1 - \frac{a
  \lambda_l T}{\sqrt{K}} \nonumber \\ && - \frac{d k_{\rm B} T}{2
  \sqrt{\pi} N K} \sum_{m=2}^{\infty}
  \frac{(-1)^{m+1}\lambda_l^{m-1}}{2^{\frac{3 m + 1}{2}}} \left(1 -
  \frac{2}{m} \right) l^{m-2} \left(
  \frac{K}{B}\right)^{\frac{m-1}{2}}
  \frac{\Gamma(\frac{m-1}{2})}{\Gamma(\frac{m}{2})}
  \bar{\zeta}_N(m-1), \end{eqnarray} in the vertically molten phase,
  where $T > T_l$.

The stability of the saddle point requires a minimum interlayer separation
\begin{equation}
  \label{lmin}
  l_{\rm min} = \mu^{-2} \sqrt{\frac{B}{K}} N^{\frac{1}{2 N}} \exp
  \left( \frac{4 \pi \kappa}{d k_{\rm B} T_{\rm c}}\right)
\end{equation}
below which the stack becomes unstable.  $l_{\rm min}$ is inversely
proportional to de Genne's penetration depth $\sqrt{K/B}$
\cite{DGbook}, which is of the order of the interlayer separation in
smectic liquid crystals.

\begin{figure}[t]
\begin{center}
\epsfxsize=6.cm \mbox{\epsfbox{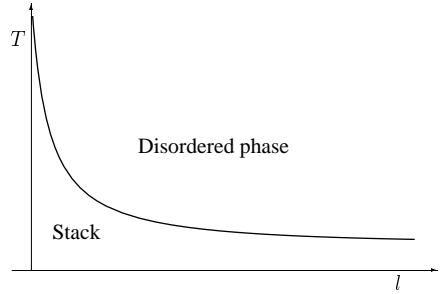}}
\end{center}
\caption{Qualitative phase diagram in the $l \times T$ plane. The
critical line is plotted for $l > l_{\rm min}$, for a fixed number
$N+1$ of membranes in the stack.  As $l$ increases, the critical
temperature $T_l$ goes asymptotically to zero.}
\label{fig:phase}
\end{figure}

The phase diagram of the stack is depicted in Fig.\ \ref{fig:phase}

\subsection{Properties of the phases}

Let us now characterize both phases in more detail.
For temperatures higher than $T_l$, the solution of the saddle
point is $\lambda = \lambda_l$.This corresponds to the disordered
phase, where the stack melts.  The normals to the membranes are
uncorrelated beyond a length scale $\lambda_l^{-1/2}$, as can be derived
from the expression for the orientational correlation function, which
in the limit $N \to \infty, l \to 0$, with constant $N l = L_{\parallel}$, reads:
\begin{equation} \label{exponential}
\langle \partial_i u({\bf x}_\perp,z) \partial_j u({\bf x}_\perp',z)
\rangle \sim \delta_{i j} {\rm e}^{-\sqrt{\lambda_l}|{\bf x}_\perp -
{\bf x}_\perp'|}.
\end{equation}
The length scale $\lambda_l^{-1/2}$ may thus be identified with the
persistence length $\xi_{\rm p}$ \cite{leibler}.

In the low-temperature phase, the solution of the saddle point is
$\lambda = 0$.  This corresponds to the ordered lamellar phase as can
be seen by examining the orientational correlation function in the
planes of the membranes.  We find in the limit $N \to \infty, l \to
0$, with constant $N l = L_{\parallel}$:
\begin{equation} \label{algebraic}
\langle \partial_i u({\bf x}_\perp,z) \partial_j u({\bf x}_\perp',z)
\rangle \sim \frac{\delta_{i j}}{|{\bf x}_\perp - {\bf x}_\perp'|^3}.
\end{equation}
This slow, algebraic fall-off of the correlation function implies that,
at large distances, the normal vectors to the membranes are still
parallel, so that the surfaces remain flat on the average.  The effect
of thermal fluctuations is suppressed, and they do not disorder the
stack.

Let us now calculate the entropy loss in the ordered phase.  By a
simple scaling argument \cite{helfstack}, this quantity should be
inversely proportional to the quadratic interlayer spacing,
\begin{equation}
  \label{deltaS}
  - T \Delta S = \frac{a}{l^2}, 
\end{equation} 
where $a$ is a temperature dependent proportionality constant. As we
shall see, as the interlayer distance increases, logarithmic
corrections must be added to (\ref{deltaS}).

The entropy loss in the ordered phase can be computed by calculating
the difference between the free energy density of a single, isolated membrane,
an the free energy density of the stack.  For small values of $\lambda$, it is
given by

\begin{equation}
  \label{deltaf} - T \Delta S = \frac{1}{2}
\sqrt{\frac{B}{K}}\left(1 + \cot\frac{\pi}{4 N}\right)\frac{1}{N l^2}
 - \frac{\lambda}{2 \pi l}\left( 1 - \ln \frac{\lambda l}{2} \sqrt{\frac{K}{B}} \right)
+ \frac{\lambda}{2 \pi N l} \sum_{n=1}^N \ln \sin
\frac{n \pi}{2 N}.
\end{equation}

Strictly speaking, the ordered phase corresponds to $\lambda=0$. In
that case, (\ref{deltaf}) agrees with (\ref{deltaS}), and we see no
correction to the entropy loss.  However, if the size of the membranes
in the stack is smaller than the characteristic length $L_{\rm p}$, an
ordered phase still exists for small values of $\lambda$ [see
discussion after Eq.\ (\ref{Lp})], in which case the corrections to
the first term in (\ref{deltaf}) will appear.

\section{Conclusions} \label{concl}

In the limit of large embedding dimension $d$, we have shown that a
stack of tensionless and incompressible membranes melts vertically
upon approaching a critical temperature, where the lamellar phase goes
over into a disordered phase.  In contrast to the low-temperature
ordered phase, where the decay of orientational correlations is
power-like, the high-temperature disordered phase is cha\-racteri\-zed
by an exponential decay of orientational correlations, with different
length scales in the transversal and longitudinal directions.

\end{document}